\documentclass[prd,aps,twocolumn,a4paper,showkeys,nofootinbib]{revtex4-1}

\newif\ifusesec
\usesectrue % or
\usepackage{graphicx,psfrag}
\usepackage{mathrsfs}
\usepackage{amsmath,amsfonts,amssymb}
\usepackage{multirow}
\usepackage{comment}
\usepackage{bm}

\DeclareSymbolFontAlphabet{\mathrsfs}{rsfs}
\DeclareMathAlphabet\mathbfcal{OMS}{cmsy}{b}{n}

\newcommand{\be}{\begin{equation}}
\newcommand{\ee}{\end{equation}}
\newcommand{\bea}{\begin{eqnarray}}
\newcommand{\eea}{\end{eqnarray}}
\newcommand{\bel}{\begin{align}}
\newcommand{\eel}{\end{align}}

\def\p{\partial}

\def\GMc2{G M_{\odot} c^{-2}}

\def\F{{\cal F}}
\def\lm{{\ell m}}
\def\teobLR{{t^{\rm EOB}_{\Omega \, \rm peak}}}
\def\teobpo{{t^{\rm EOB}_{\Omega_{\rm orb}^{\rm max}}}}

\def\tAmax{{t_{A_{22}^{\rm max}}^{\rm NR}}}
\def\tnrextr{{t_{\rm extr}^{\rm NR}}}
\def\teobNQC{{t^{\rm EOB}_{\rm NQC}}}
\def\lm{{\ell m}}
\def\p{\partial}
\def\v{v_\varphi}

\def\de{\partial}
\def\lm{{\ell m}}

\def\ii{{\rm i}}

\def\F{{\cal F}}

\def\hr{\hat{r}}
\def\ha{{\hat{a}}}

\DeclareSymbolFontAlphabet{\mathrsfs}{rsfs}
\DeclareMathAlphabet{\mathcal}{OMS}{cmsy}{m}{n}

\usepackage{color}

\definecolor{cyan}{rgb}{0,0.9,0.9}
\definecolor{orange}{rgb}{0.9,0.5,0}
\definecolor{magenta}{rgb}{1,0,1}
\definecolor{purple}{rgb}{0.8,0.4,0.8}
\definecolor{gray}{rgb}{0.8242,0.8242,0.8242}

\begin{document}

\title{A new effective-one-body description of\\
 coalescing nonprecessing spinning  black-hole binaries}

\author{Thibault \surname{Damour}}
\author{Alessandro \surname{Nagar}}

\affiliation{$^1$Institut des Hautes Etudes Scientifiques, 91440 Bures-sur-Yvette, France}

\date{\today}

\begin{abstract}
We present a new, tunable effective-one-body (EOB) model of the motion and radiation of coalescing black hole binaries
with arbitrary mass ratio and aligned spins. The most novel feature of our formalism is the introduction, and systematic
use, of the (gauge-invariant) concept of {\it centrifugal radius} $r_{c}$. In the spinning small mass-ratio limit, the main
radial potential expressed in terms of $r_{c}$ differs very little (and only multiplicatively so) from the usual Schwarzschild
potential $1-2M/r_{c}$. This motivates a new, multiplicative way of blending finite-mass-ratio deformations with spin-deformations.
In the present exploratory work we consider a minimal version of our spinning EOB model (containing essentially only
two adjustable parameters: one in the Hamiltonian and one in the waveform) and calibrate its (dominant mode) waveform
against a sample of fifteen equal-mass, equal-spin waveforms produced by the SXS collaboration, and covering the dimensionless
spin range $-0.95\leq \chi \leq +0.98$. The numerical relativity / EOB phasing disagreement remains remarkably small ($\lesssim \pm 0.15$~rad) 
over the entire spin range.
\end{abstract}


\maketitle

\section{Introduction}
Coalescing black hole binaries are the most promising sources for the network of
ground-based, kilometer-size interferometric gravitational wave (GW) detectors that 
is about to come online at improved sensitivities~\cite{ligo,virgo}.
It has been pointed out long ago~\cite{Damour:2001tu} that there is a  bias favoring
the detection of  black hole systems with large, aligned spins. In view of the large parameter space of
spinning binaries $\left(m_1,m_2,{\bf S}_1,{\bf S}_2\right)$  and the need
to have in hands tens of thousands of waveform templates for detection purposes, 
it is useful to develop semi-analytic techniques to produce accurate templates.
One promising method towards this goal is the analytical effective-one-body (EOB) 
formalism~\cite{Buonanno:1998gg,Buonanno:2000ef,Damour:2000we,Damour:2008gu}. 
Several different ways of incorporating spin effects in the EOB formalism have
been explored~\cite{Damour:2001tu,Buonanno:2005xu,Damour:2008qf,Barausse:2009xi,
Barausse:2009aa,Barausse:2009xi,Nagar:2011fx,Barausse:2011ys,Balmelli:2013zna}.

Thanks to recent advances in Numerical Relativity (NR)(see e.g. Refs.~\cite{Mroue:2013xna,Hinder:2013oqa}), 
it has been possible, over the last years, to incorporate crucial nonperturbative
information in the EOB formalism so as to produce improved ``EOBNR'' waveforms.
This has been done both for nonspinning EOB models~\cite{Damour:2002qh,Buonanno:2006ui,Buonanno:2007pf,
Damour:2007yf,Damour:2007vq,Damour:2008te,Damour:2009kr,Buonanno:2009qa,Baiotti:2010xh,Baiotti:2011am,
Damour:2011fu,Bernuzzi:2012ci,Pan:2011gk,Pan:2013tva} and for spinning
(precessing) configurations~\cite{Pan:2009wj,Pan:2013rra,Taracchini:2012ig,Taracchini:2013rva}.
Progress in gravitational self-force theory has also allowed one to acquire useful strong-field
information~\cite{Detweiler:2008ft,Blanchet:2009sd,Barack:2009ey,Barack:2010ny,Barausse:2011dq,
Akcay:2012ea,Shah:2012gu,Dolan:2013roa,Bini:2014ica}. 

Here, we shall propose a new way
of defining an EOB Hamiltonian containing spin-orbit and spin-spin 
interactions. In addition, we will calibrate our new spinning EOB (SEOB) model
by using a sample of Caltech-Cornell Simulating eXtreme Spacetime 
(SXS)~\cite{SXS:catalog,Chu:2009md,Lovelace:2010ne,Lovelace:2011nu,Buchman:2012dw,Mroue:2012kv,Mroue:2013xna,Hemberger:2013hsa}
NR waveforms covering the dimensionless spin range $-0.95\leq \chi \leq 0.98$.
As a proof of principle we shall only consider here the simplest case of equal
mass $m_1=m_2$, equal, parallel (or anti parallel) spin ($\chi_1=\chi_2=\chi$) where
$\bm{\chi}_A\equiv {\bf S}_A/m_A^2$ ($G=c=1$).
We leave to future work a more extensive comparison/calibration of our new
SEOB Hamiltonian.

\smallskip

\section{Revisiting the Kerr Hamiltonian}
\label{sec:kerr}

Here, we reexamine the structure of the Hamiltonian of a test particle 
on a Kerr background of mass $M$ and spin parameter $a=S/M$, 
to motivate our definition of a new SEOB model. 
The Hamiltonian of a spinning test particle of mass $\mu$ and spin ${\bf S}_*$
can be decomposed in ``orbital'' and ``spin-orbit'' parts:
\be
\label{eq:H_K}
H_{\rm Kerr} = H_{\rm orb}^{\rm Kerr} + H^S_{\rm so}({\bf S}) + H^{S_*}_{\rm so}({\bf S_*}).
\ee
Here the orbital Hamiltonian (which also contains all interactions that are even 
in spins such as spin-square effects $\propto S^2$) is 
\begin{widetext}
\be
\label{eq:HorbFull}
H_{\rm orb}^{\rm Kerr} = \sqrt{  A(r,\theta)\left(\mu^2 + \dfrac{\Delta(r) p_r^2}{r^2 + a^2 \cos^2\theta}
         +\dfrac{p_\theta^2}{r^2 + a^2 \cos^2\theta} + 
          \dfrac{r^2 + a^2 \cos^2\theta}{{\cal R}^4(r) + a^2\Delta(r) \cos^2\theta}\dfrac{p_\varphi^2}{\sin^2\theta}\right)},
\ee
\end{widetext}
where
\begin{align}
\Delta(r) &\equiv r^2 - 2M r + a^2\\
{\cal R}^4(r)&\equiv \left(r^2+a^2\right)^2 - a^2\Delta(r) \nonumber\\
             & = r^4 + r^2 a^2 + 2M r a^2 \\
A(r,\theta) & \equiv \frac{\Delta(r)(r^2 + a^2\cos^2\theta)}{{\cal R}^4 + a^2\Delta(r) \cos^2\theta}
\end{align}

and where the part of the spin-orbit~\footnote{Here we gather in the spin-orbit Hamiltonian all
 the effects that are odd in spins: $S^1,S^3,\ldots$.} Hamiltonian linked to the background 
spin ${\bf S}$ reads 
\be
H_{\rm so}^S({\bf S})\equiv G_S(r,\theta)  \, {\bf L}\cdot {\bf S}.
\ee
Here ${\bf L}$ denotes the orbital angular momentum of the particle, while
the gyro-gravitomagnetic function entering $H_{\rm so}^S({\bf S})$ reads
\be
\label{eq:GS}
G_S(r,\theta) \equiv \dfrac{2 r}{{\cal R}^4(r) + a^2\Delta(r) \cos^2\theta}.
\ee
Methods for computing the spin-orbit Hamiltonian  $H^{S_*}_{\rm so}({\bf S_*})$ linked to
the spin of the particle have been discussed in~\cite{Porto:2005ac,Damour:2008qf,Barausse:2009aa,Hartung:2013dza}.

The structure of this Hamiltonian can be clarified by introducing the (gauge-invariant) 
concept of ``centrifugal radius'' $r_c(r)$, defined so that the orbital part of the 
Hamiltonian ruling equatorial orbits ($\theta=\pi/2$) can be written as
\be
\label{eq:HorbEq}
H_{\rm orb,eq}^{\rm Kerr}(r,p_r,p_\varphi) = \sqrt{A^{\rm eq}(r)\left(\mu^2 + \dfrac{p_\varphi^2}{r_c^2}
+\dfrac{p_r^2}{B^{\rm eq}(r)}\right)},
\ee
with the usual, relativistic centrifugal energy term $ \mu^2 + p_\varphi^2/r_c^2$.
By comparing with Eq.~\eqref{eq:HorbFull} one obtains
\be
r_c^2 \equiv \frac{{\cal R}^4(r)}{r^2} = r^2 + a^2 + \dfrac{2M a^2}{r}.
\ee
In addition, we have 
\be
\label{eq:AK_eq}
A_{\rm eq}(r)\equiv \dfrac{\Delta(r)}{r_c^2}=\left(1-\dfrac{2M}{r_c}\right)\dfrac{1+\dfrac{2M}{r_c}}{1+\dfrac{2M}{r}},
\ee
and $B^{\rm eq}(r)=r^2/\Delta$ so that 
\be
A^{\rm eq}(r) B^{\rm eq}(r) = \dfrac{r^2}{r_c^2}.
\ee
We also note that
\be
G_{S}^{\rm eq}(r_c)=\dfrac{2}{r r_c^2}.
\ee
Equation~\eqref{eq:AK_eq} displays a remarkable fact, which is rather hidden in usual formulations of
the Kerr Hamiltonian: while the usual (gauge-dependent) Boyer-Lindquist radius of the outer horizon
($\Delta(r_H^+)=0$), which is also the location where the main radial $A$ potential entering the 
equatorial dynamics has a zero, strongly depends on the spin parameter $a$, $r_H^+=M+\sqrt{M^2-a^2}$,
the corresponding (gauge-invariant) value of the centrifugal radius $r_c$ does not depend on $a$ and
is simply equal to the usual Schwarzshild-coordinate value, $r_c=2M$. 
In other words, when using $r_c$ as radial coordinate, the crucial $A$ potential factorizes 
in the product
\be
\label{A_vs_rc}
A_{\rm eq}(r_c) = A_{\rm Schw}(r_c)\hat{A}(r_c)
\ee
of the usual Schwarzschild $A_{\rm Schw}(r_c)=1-2M/r_c$ and of a correcting factor $\hat{A}$ 
\begin{align}
\label{eq:hatArc}
\hat{A}(r_c)   &     = \dfrac{1+\dfrac{2M}{r_c}}{1+\dfrac{2M}{r}}\nonumber\\
               &\approx \dfrac{1}{1 + \dfrac{M a^2}{r_c^3} 
                              + \dfrac{3}{4}\dfrac{M a^4}{r_c^5} 
                              + \dfrac{5}{2}\dfrac{M^2 a^4}{r_c^6} 
                              + {\cal O}\left(\dfrac{1}{r_c^7}\right)}
\end{align}
This natural factorization of the Kerr $A$ potential will be the crucial motivation for the definition
of our new SEOB formalism below.

The correcting factor $\hat A(r_c)$ embodies the multipolar structure of the Kerr hole, e.g. at lowest 
order in a PN expansion one finds
\begin{align}
A_{\rm PN}(r_c) &= \left(1-\dfrac{2M}{r_c}\right)\left(1-\dfrac{Ma^2}{r_c^3}+\dots\right) \nonumber\\
             &\approx 1 - \dfrac{2M}{r_c} - \dfrac{Ma^2}{r_c^3},
\end{align}
where $-\dfrac{Ma^2}{r_c^3}$ is the quadrupole gravitational potential term (seen in the equatorial plane).

\begin{comment}
Point: when expressed in terms of $r_c$ $a^2$ never appears alone, but only multiplied by a 
power of $M$, so that
when $M$ goes to zero all spin effects disappear (which is not true for the Boyer-Lindquist case, 
spheroidal coordinate system).
Essentially Kerr=Schwarzschild plus some other effects: there is a multiplicative modifications 
of the A potential by a factor which does not change the location of the horizon, but introduces 
extra attractive forces lined to the quadrupole
and higher moments of Kerr. Figure of the $A(u_c)$ potential.
 $Ma^2$ is the value of the quadrupole.
Not $a^2$ has physical meaning. 
\end{comment}

\begin{figure}[t]
\begin{center}
 \includegraphics[width=0.45\textwidth]{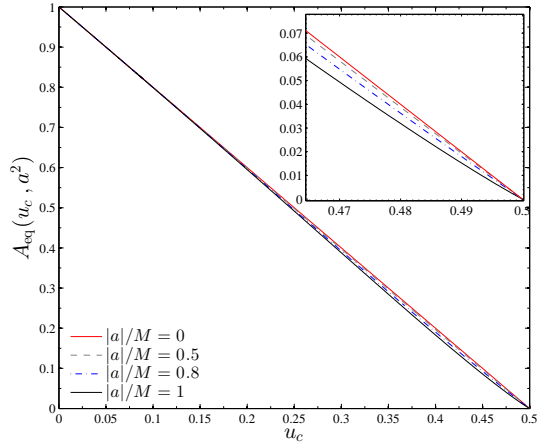}
    \caption{ \label{fig:kerrA} Weak dependence on $a^2$ of the equatorial 
    Kerr radial potential $A_{\rm eq}$, Eq.~\eqref{eq:AK_eq}, as a function of 
    $u_c=M/r_c$, up to the horizon $u_c^H=1/2$.} 
\end{center}
\end{figure}

The (multipolar) correcting factor $\hat{A}(r_c)$ is everywhere smaller than one and
larger than $2/3$, a value reached only at the horizon $r_c=2M$ and for maximum spin $a=M$.
The derivative of the function $r_c(r)$ is equal to
\be
\dfrac{d r_c}{dr} = \dfrac{r}{r_c}\left(1-\dfrac{Ma^2}{r^3}\right)
\ee
so that $r_c(r)$ reaches a minimum at $r=r_{\rm min}=(Ma^2)^{1/3}$. One sees that, when $a<M$,
$r_{\rm min}<M$ so that $r_c^{\rm min}< 2M$. However, when $a\to M$, $r_{\rm min}\to M$ and
$r_c^{\rm min}\to 2M$. The inverse function expressing the original Boyer-Lindquist 
radial coordinate $r$ in terms of the centrifugal radius $r_c$ is 
well defined on the interval $r_c\geq 2M$, i.e. outside the horizon, and reads
\be
r = 2\sqrt{ \dfrac{ r_c^2-a^2 }{3}} \cos\left( \dfrac{\pi}{6}+\frac13 \arcsin \dfrac{M a^2}{\left(\dfrac{r_c^2-a^2}{3}\right)^{3/2} }\right).
\ee
The multiplicative modification of $A(r_c)$ by a factor, depending on $a^2$, 
that does not change the location of the horizon, but introduces extra attractive 
forces linked to the quadrupole and higher moments of Kerr is illustrated in Fig.~\ref{fig:kerrA}.
It is striking how small is the modification of the Kerr $A$ potential due to the 
$a^2$ terms when plotted as a function of the inverse centrifugal 
radius~\footnote{By contrast, let us recall that the usual Boyer-Lindquist location
of the horizon varies between 2 $M$ (for $a=0$) and $1M$ (for $a=M$).} $u_c=M/r_c$.
Numerically, the difference $A(u_c\;,a)-A(u_c\;,0)$ (which vanishes both at $u_c=0$
and $u_c=\frac12$)
reaches, around $u_c\approx 0.4$,
an extremum approximately equal to $-0.004$ for $|a|=0.5 M$, to $-0.01$ for $|a|=0.8 M$ and 
to $-0.016$ for $a=1M$. This indicates that the most important physical effects determining
the energetics of circular orbits  comes from the interplay between  $A_{\rm Schw}(r_c)=1-2M/r_c$
and the spin-orbit coupling ($G_S(r_c) L S$), with only a relatively small contribution of
(quadrupolar) spin-squared effects.

Similar multiplicative modifications occur in the (equatorial) gyro-gravitomagnetic 
function which reads
\be
\label{GSeq}
G_{S}^{\rm eq}(r_c)=\dfrac{2}{r r_c^2}=\dfrac{2}{r_c^3}\hat{G}_S(r_c),
\ee
leading to a spin-orbit coupling
\be
H_{\rm so,eq}^{\rm S} = \dfrac{2}{r_c^3}\hat{G}_S M a L , %+  \sqrt{\left(1-\dfrac{2M}{r_c}\right)\hat{A}\left(\mu^2 + \dfrac{p_\varphi^2}{r_c^2}}}
\ee
where
\be
\hat{G}_S(r_c ) = \dfrac{r_c}{r} \approx 1 + \frac12 \dfrac{a^2}{r_c^2} + \dfrac{M a^2}{r_c^3} + {\cal O}\left(\dfrac{1}{r_c^4}\right).
\ee
Finally, let us note that, when considering general nonequatorial orbits all the functions appearing in
the Hamiltonian naturally factorize as the product of their equatorial (radial-dependent) value with
a $\cos\theta$-dressing factor of the type
\be
\label{eq:fcosth}
\dfrac{1+f_1(r)\cos^2\theta}{1+f_2(r)\cos^2\theta}.
\ee
For instance, the $A(r,\theta)$ function reads
\be
\label{eq:AK_rth}
 A(r,\theta) = A_{\rm eq}(r)
\dfrac{1 + \dfrac{a^2 \cos^2\theta}{r^2}}{1+ \dfrac{a^2\Delta\cos^2\theta }{r^2 r_c^2}}.
\ee

%===================================
% Fig.1: phasing in presence of spin
%===================================
%-----------------

%varies between 
%$2/3\hat{A}<1$
%in the maximum spin and at the horizon.

%solution of the cubic equation when solving $r$ in terms of $r_c$.

%==========================================
\section{Dynamics: A new SEOB Hamiltonian }
\label{sec:EOB_dyn}
%==========================================
Some of the main challenges in previous definitions of SEOB models
were: (i) blending Kerr spin effects in 
$\Delta(r)=r^2-2Mr + a^2=r^2\left(1-2M/r + a^2/r^2\right)$ with 
finite-mass-ratio deformations  (depending on $\nu\equiv m_1 m_2/(m_1+m_2)^2$)
in $A= 1-2 M/r + 2\nu (M/r)^3 + \dots$; and (ii) resumming spin-orbit 
and spin-spin effects. 
A first suggestion~\cite{Damour:2001tu} to address issue (i) was to deform 
the Kerr $\Delta(r)$ function (considered in Boyer-Lindquist-type coordinates) 
into the Pad\'e-resummed function 
$\Delta_t(r)=r^2 P^1_n[1-2u+a^2 u^2 + 2\nu u^3+\dots]$ where $u=M/r$.
More recently, it was suggested~\cite{Barausse:2009xi} to enforce the existence
of two zeros in $\Delta_t(r)$ of the form
$r_{H^\pm}^{\rm EOB}=\left(M\pm \sqrt{M^2-a^2}\right)(1-K\nu)$ 
(where $K$ is an adjustable parameter) by factoring $\Delta_t(r)$ as
\begin{align}
&\Delta_t(r) = \dfrac{r^2 a^2}{M^2}\left(u-\dfrac{M}{r_{H^+}^{\rm EOB}}\right)
\left(u-\dfrac{M}{r_{H^-}^{\rm EOB}}\right)\nonumber\\
&\times \left[1+\nu\Delta_0 + \nu\ln(1+\Delta_1 u+\Delta_2 u^2 + \Delta_3 u^3 + \Delta_4 u^4)\right]\nonumber.
\end{align}
Here, we introduce a novel (and more natural) way of blending the $a^2$-deformation effects with the
$\nu$-deformation ones based on the new way of considering the Kerr $A$-potential
explained in the previous section. Namely, we shall extend the definition of a
{\it centrifugal radius} $r_c$ to the spinning, comparable-mass case and decompose
its (equatorial) radial $A$-potential as the product of the usual EOB, $\nu$-deformed, 
nonspinning (Pad\'e-resummed) $A$-potential $A_{\rm orb}(u_c,\nu)$, expressed as a 
function of $u_c\equiv M/r_c$, by a $\nu$-deformed version of the multipolar correcting 
factor $\hat{A}(r_c)$, Eq.~\eqref{eq:hatArc}.

Concerning the challenge of resumming spin-orbit effects, it was previously addressed
in Refs.~\cite{Damour:2008qf,Barausse:2009xi,Nagar:2011fx,Barausse:2011ys}. Here, we shall
first follow the Damour-Jaranowski-Sch\"afer (DJS)~\cite{Damour:2000we,Damour:2008qf}
philosophy of working in a gauge that eliminates the ${\bf p}^2$ dependence to keep only
a dependence on $1/r$ and $p_{r_*}^2$. In addition, as previous 
work~\cite{Damour:2008qf,Barausse:2009xi,Nagar:2011fx,Barausse:2011ys,Taracchini:2012ig,Balmelli:2013zna}
has shown the importance of controlling the natural taming of the gyrogravitomagnetic couplings
by higher-order PN corrections, we shall resum them by means of $P^0_m$ (i.e., inverse Taylor) Pad\'e approximants.

In the present work, we limit ourselves, for simplicity and orientation, to considering nonprecessing (parallel)
spins. This means that we shall define a comparable-mass deformation of the equatorial-Hamiltonian-dynamics of
Sec.~\ref{sec:kerr}, leaving to future work the task of $\nu$-deforming the various $\cos\theta$ dressing factors 
of Eq.~\eqref{eq:fcosth}.
We recall that the full EOB Hamiltonian $H$ is expressed in terms of the effective Hamiltonian $H_{\rm eff}$ (which 
generalizes the Hamiltonian of a test-particle) as
\be
H = M\sqrt{1+2\nu\left(\dfrac{H_{\rm eff}}{\mu}-1\right)}.
\ee
Here and henceforth, $M\equiv m_1+m_2$, $\mu=m_1 m_2/M$ and $\nu=\mu/M$, by contrast with the previous 
section where $M$ denoted the large mass and $\mu$ the small one.
We decompose the effective Hamiltonian $H_{\rm eff}$ in an orbital part $H_{\rm orb}^{\rm eff}$ 
(which gathers all the terms that are even in spins) and in two spin-orbit pieces 
(that are odd in spins):
\be
\label{eq:Heff}
H_{\rm eff} =  H^{\rm eff}_{\rm orb} + G_S {\bf L}\cdot {\bf S} + G_{S_*}{\bf L}\cdot {\bf S}_* ,
\ee
where
\be
\label{eq:Horb}
H_{\rm orb}^{\rm eff} = \sqrt{p_{r_*}^2 + A(r,\nu,S_1,S_2)\left(\mu^2 + \dfrac{L^2}{r_c^2} + Q_4\right)}.
\ee
Here, ${\bf L}={\bf r}\times {\bf p}=L \,{\bf e}_z$ denotes the orbital angular 
momentum (with modulus $L=p_\varphi$) and  the comparable-mass analogs of the 
background and (rescaled) test spins $({\bf S},{\bf S}_*)$
are, following Ref.~\cite{Damour:2008qf}, defined as
\begin{align}
{\bf S}   &={\bf S}_1 + {\bf S}_2, \\
{\bf S_*} &=\dfrac{m_2}{m_1}{\bf S}_1+\dfrac{m_1}{m_2}{\bf S}_2.
\end{align}
The building blocks, $A,r_c^2,Q_4,G_S,G_{S_*}$ entering the effective 
Hamiltonian~\eqref{eq:Heff} are defined next.

\subsection{Effective orbital Hamiltonian}
Henceforth we will often (but not always~\footnote{The context will make it clear whether rescaled variables are used or not.}) 
work with rescaled variables: $r\equiv r^{\rm phys}/M$, ${\bf p}\equiv {\bf p}^{\rm phys}/\mu$, $p_\phi\equiv L\equiv L^{\rm phys}/(\mu M)$ as
well as $H\equiv H^{\rm phys}/M$ and $\hat{H}_{\rm eff}\equiv H_{\rm eff}^{\rm phys}/\mu$.
We also introduce $u=M/r^{\rm phys}=1/r$ and   $u_c=M/r_c^{\rm phys}=1/r_c$. 
The 5PNlog-accurate PN expansion of the orbital (nonspinning) part of the $A$ potential,
considered as a function of $r_c$, reads
\begin{widetext}
\begin{align}
\label{AorbPN}
&A_{\rm orb}^{PN}(u_c)   = 1 - 2 u_c+2\nu u_c^3 + \nu a_4 u_c^4 + \nu (a_5^c + a_5^{\rm \log} \ln u_c)u_c^5 + \nu (a_6^c + a_6^{\rm\log} \ln u_c)u_c^6 ,
\end{align}
\end{widetext}
where $a_4=94/3 - (41/32)\pi^2$~\cite{Damour:2000we,Damour:2001bu} and where we shall use the analytically 
known logarithmic coefficients~\cite{Damour:2009sm,Blanchet:2010zd,Barack:2010ny,Barausse:2011dq}
\be
a_5^{\log}     =  \frac{64}{5} , \qquad
a_6^{\log}(\nu)  =  -\frac{7004}{105} - \frac{144}{5}\nu .
\ee
As in Ref.~\cite{Damour:2012ky}, we shall (phenomenologically) fix $a_5^c$ to the fiducial
value $a_5^c=23.5$~\cite{Barausse:2011dq,Akcay:2012ea,Bini:2013zaa} and use $a_6^c$ as our unique,
orbital-Hamiltonian adjustable effective parameter, whose NR calibration will be discussed 
below. We Pad\'e resum~\eqref{AorbPN} as
\be
A_{\rm orb}(u_c,\nu) = P^1_5[A_{\rm orb}^{\rm PN}](u_c).
\ee
The complete equatorial $A$ function entering $\hat{H}^{\rm eff}_{\rm orb}$ is then defined,
on the model of the Kerr $A$ function~\eqref{A_vs_rc}, as the product 
\be
\label{A_orb}
A(r,\nu,S_1,S_2) = \left[A_{\rm orb}(u_c,\nu)\dfrac{1+ 2 u_c}{1+ 2 u}\right]_{u_c(u,S_1,S_2)}
\ee
where the comparable-mass version of the spin-dependent function $u_c(u)$ relating $u_c$ 
to the Boyer-Lindquist type variable $u=1/r$ will be introduced below. As indicated here, though we conceptually
view $A$ as a function of $r_c$, it will be convenient to express the Hamiltonian dynamics 
in terms of a Boyer-Lindquist type radius $r$, and, as usual, of a related tortoise radial
momentum $p_{r_*}$.
The other building elements of $H_{\rm eff}$ are defined through
\be
D \equiv A\,B = \dfrac{r^2}{r_c^2}D_{\rm orb}(u_c),
\ee
with 
\be
D_{\rm orb}(u_c) = \dfrac{1}{1 + 6\nu u_c^2 + 2(26 - 3\nu)\nu u_c^3},
\ee
(note that the coefficient of $\nu u_c^3$ here was misprinted as
$2(23-3\nu)$ in Ref.~\cite{Damour:2012ky})
and
\begin{align}
p_{r_*}^2  & = \dfrac{A}{B} p_r^2,\\
Q_4      & =  2\nu(4-3\nu) p_{r_*}^4 u_c^2.
\end{align}

\subsection{Spin-orbit interaction terms}
On the model of the multiplicative structure~\eqref{GSeq} of the gyrogravitomagnetic coupling
$G_S$ entering the equatorial Kerr Hamiltonian, we write 
\begin{align}
G_S    &= G_S^0 \hat{G}_S,\\
G_{S_*} &= G_{S_*}^0 \hat{G}_{S_*},\\
\end{align}
where
\begin{align}
G_S^0 &= 2 u u_c^2,\\
G_{S_*}^0 &= \dfrac{3}{2} u_c^3.
\end{align}
Here, $(\hat{G}_{S},\hat{G}_{S_*})$ are PN correcting factors $\hat{G}_S=1 + {\cal O}(1/c^2)$, 
$\hat{G}_{S_*}=1 + {\cal O}(1/c^2)$. We take them in DJS gauge (i.e., as function of $1/r_c$
and $p_{r_*}^2$ only), resumming their NNLO PN expansion~\cite{Nagar:2011fx,Barausse:2011ys} 
by means of inverse Taylor, $P^0_m$, Pad\'e approximants.
In addition, we include two more $\nu=0$ terms in $\hat{G}_{S_*}$, coming from the 
circular limit of a spinning test-particle in a Schwarzschild 
background~\cite{Barausse:2009aa,Bini:2014ica}
\begin{align}
\label{eq:GSstest}
&\hat{G}_{S_*}^{\rm circ}(\nu=0)=\dfrac{2}{1+\dfrac{1}{\sqrt{1-3u_c}}}\nonumber\\
&\approx \left(1+ \dfrac{3}{4}u_c + \dfrac{27}{16}u_c^2 + \dfrac{135}{32}u_c^3 + \dfrac{2835}{256} u_c^4+\dots\right)^{-1}.
\end{align}
Their explicit expressions read 
\begin{align}
\hat{G}_{S}  & = \bigg(1  + c_{10} u_c + c_{20} u_c^2 + c_{30} u_c^3 \nonumber\\
            & \hspace{1cm}+ c_{02} p_{r_*}^2 + c_{12} u_c p_{r_*}^2 + c_{04} p_{r_*}^4\bigg)^{-1},\\
\hat{G}_{S_*} & = \bigg(1  + c_{10}^* u_c + c_{20}^* u_c^2 + c_{30}^* u_c^3 +  c_{40}^* u_c^4 \nonumber\\
             &+ c_{02}^* p_{r_*}^2 + c_{12}^* u_c p_{r_*}^2 + c_{04}^* p_{r_*}^4\bigg)^{-1} .             
\end{align}
Starting from Eqs.~(55) and~(56) of Ref.~\cite{Nagar:2011fx,PhysRevD.88.089901} 
and taking their $P^0_m$ approximants one explicitly obtains
\begin{align}
c_{10} & =\dfrac{5}{16}\nu,\\
c_{20} & = \dfrac{51}{8}\nu + \dfrac{41}{256}\nu^2,\\
\label{eq:c3}
c_{30} & = \nu c_3,\\
c_{02} & = \dfrac{27}{16}\nu, \\
c_{12} & = 12\nu - \dfrac{49}{128}\nu^2, \\
c_{04} & = -\dfrac{5}{16}\nu + \dfrac{169}{256}\nu^2,\\
c_{10}^* & =\dfrac{3}{4}   + \dfrac{\nu}{2}.\\
c_{20}^* & =\dfrac{27}{16} + \dfrac{29}{4}\nu + \dfrac{3}{8}\nu^2, \\
c_{02}^* & =\dfrac{5}{4}   + \dfrac{3}{2}\nu, \\
\label{eq:c3s}
c_{30}^* & =\dfrac{135}{32} + \nu c_3,\\
c_{40}^* & = \dfrac{2835}{256},\\
c_{12}^* & =4 + 11\nu-\dfrac{7}{8}\nu^2,\\
c_{04}^* & =\dfrac{5}{48}  + \dfrac{25}{12}\nu + \dfrac{3}{8}\nu^2. 
\end{align}
We included  $\nu$-dependent tunable next-to-next-to-next-to-leading-order (NNNLO) 
contributions {\it both} in $(\hat{G}_{S},\hat{G}_{S_*})$. They are taken here as 
being simply proportional to $u_c^3$. In this exploratory  investigation we shall conflate 
these two (apriori independent) terms in a common NNNLO contribution parametrized 
by the single parameter $c_3$ entering both $c_{30}$ ad $c_{30}^*$, 
Eqs.~\eqref{eq:c3},~\eqref{eq:c3s}. Note that 
in $c_{30}^*$ (and $c_{40}^*$) we included also the $\nu$-independent 
(spinning test-particle) contribution coming from Eq.~\eqref{eq:GSstest}.
Recently, combined progress in numerical and analytical gravitational
self force calculations of spin-orbit effects~\cite{Dolan:2013roa,Bini:2014ica}
has brought an improved knowledge of $G_{S_*}$. We leave to future work the
incorporation of this knowledge in the SEOB model.

\subsection{Spin-spin interaction}
We have seen above that in the Kerr Hamiltonian expressed in $r_c$ coordinates 
the quadrupole moment of the Kerr black hole was modifying the main radial potential 
by $\delta A^{\rm quadrupole} = - M a_{\rm Kerr}^2/r_c^3$.
It was shown in~\cite{Damour:2001tu} that, to leading PN order, the combination of 
quadrupolar $S_1^2, S_2^2$ effects with  spin-spin $S_1 S_2$ effects, could be
incorporated simply by changing the Kerr spin parameter $a$ entering the 
Kerr Hamiltonian by the leading-order effective Kerr 
parameter\footnote{As shown in Eq.~(2.55) 
of~\cite{Damour:2001tu} this is also true at the vectorial level.}
\be
\label{eq:a1a2}
a \equiv a_1 + a_2= \dfrac{S_1}{m_1} + \dfrac{S_2}{m_2} = \dfrac{S+S_*}{m_1 + m_2}.
\ee
In other words, we can include the leading order effect of spin-spin interactions
by defining the functional link between the SEOB centrifugal radius $r_c$ and
its Boyer-Lindquist type analog $r$ by the relation
\be
\label{eq:rcLO}
\left[r_c^{\rm LO}(r,S_1,S_2)\right]^2 = r^2 + a^2 + 2\dfrac{M a^2}{r},
\ee
where, now, $M=m_1 + m_2$ and $a$ is given by Eq.~\eqref{eq:a1a2}. 

At the next-to-leading (NLO) order in spin-spin effects we need to correct the quadrupolar
and spin-spin interaction terms $\sim  M S^2/r_c^3$ (viewed in the equatorial plane) 
by adding PN corrections $\sim  \nu S^2/r_c^3 \left(p^2 + p_r^2 + \dfrac{1}{r}\right)$.
Starting from the EOB reformulation~\cite{Balmelli:2013zna} of the PN-expanded NLO 
spin-spin results of Refs.~\cite{Steinhoff:2007mb,Steinhoff:2008ji},
one finds that, when including for simplicity only the circular, equatorial (spin-aligned) contributions, 
we can take into account NLO spin-spin effects by simply modifying the definition of the
function $r_c^2(r)$ as
\be
\label{eq:rcNLO}
\left[r_c^{\rm NLO}(r,S_1,S_2)\right]^2 = r^2 + a^2 + 2\dfrac{M a^2}{r}+\delta a^2(r),
\ee
where $\delta a^2(r)$ is 
\be
\label{eq:del_a}
\delta a^2(r)=  \dfrac{M^3}{r}\bigg\{(a_{11}+c_{11})\chi_1^2 + (a_{22}+c_{22})\chi_2^2 
+ (a_{12}+c_{12})\chi_1\chi_2\bigg\}
\ee
with $\chi_A=a_A/m_A=S_A/m_A^2$ and~\cite{Balmelli:2013zna} 
\begin{widetext}
\begin{align}
a_{11} &= \dfrac{\nu}{16}\left[-32\nu - 22\nu^2 + X_{1/2}(21-44\nu-44\nu^2) + (X_{1/2})^2(-21\nu-22\nu^2)\right],\\
a_{12} &= \dfrac{\nu}{8}\left[24 - 53\nu-44\nu^2 + (X_{1/2} + X_{2/1})(-32\nu-22\nu^2)\right],\\
a_{22} &= \dfrac{\nu}{16}\left[-32\nu - 22\nu^2 + X_{2/1}(21-44\nu-44\nu^2) + (X_{2/1})^2(-21\nu-22\nu^2)\right],\\
c_{11} &= \dfrac{\nu}{16}\left[88\nu + 14\nu^2 + X_{1/2}(-117+196\nu+28\nu^2) + (X_{1/2})^2(117\nu+14\nu^2)\right],\\
c_{12} &= \frac{\nu}{8}\left[-120+229\nu+28\nu^2 + (X_{1/2} + X_{2/1})(112\nu + 14\nu^2)\right],\\
c_{22} &= \frac{\nu}{16}\left[88\nu + 14\nu^2 + X_{2/1}(-117+196\nu+28\nu^2) + (X_{2/1})^2(117\nu+14\nu^2)\right].
\end{align}
\end{widetext}
Here
\be
X_{1/2}\equiv \frac{X_1}{X_2} \qquad X_{2/1}\equiv \frac{X_2}{X_1},
\ee
with
\be
X_1 \equiv \dfrac{m_1}{M} = \frac12(1+\sqrt{1-4\nu}), \qquad X_2\equiv \dfrac{m_2}{M}  = 1-X_1,
\ee
where we use the convention $m_1\geq m_2$. 

In the equal-mass, equal-spin case, $m_1=m_2$,
$\chi_1= \chi_2=\chi$ (so that $a\equiv a_1+a_2=M\chi$), Eq.~\eqref{eq:del_a} gives
$\delta a^2 = -\dfrac{9}{8}\dfrac{M a^2}{r}$ which amounts
to replacing the LO term $2\dfrac{M a^2}{r}$ in
Eq.~\eqref{eq:rcLO} by $\dfrac{7}{8}\dfrac{M a^2}{r}$.

In the present work we shall use everywhere the $H_{\rm orb}^{\rm eff}$-related 
NLO functional link $r_c(r)$ given by Eq.~\eqref{eq:rcNLO}. Note, however, that this 
use means that we are introducing specific corresponding $\nu$-dependent, 
spin-quadratic, NLO corrections to other parts of the Hamiltonian (such as $G_S,G_{S_*}$).

%==========================================================
\section{Radiative sector: radiation reaction and waveform}
\subsection{Flux and waveform at future null infinity}
%==========================================================
In absence of a robust strategy for resumming the analytically
predicted~\cite{Bini:2012ji} radial contribution to the radiation
reaction, we set it to zero, $\F_r=0$ (as we had done in most of
our previous EOB work).
As a preparation for defining the azimuthal contribution $\F_\varphi$ 
to radiation reaction, as well as the waveform, especially during the 
post-Keplerian plunge~\cite{Damour:2006tr}, we define a non-Keplerian 
``azimuthal'' velocity $v_\varphi$ as
\be
v_\varphi \equiv r_\Omega \Omega.
\ee
Here
\be
\label{eq:rOmg}
r_\Omega \equiv  \left\{\dfrac{(r_c^3\psi_c)^{-1/2}+\tilde{G}}{H}\right\}^{-2/3}_{p_{r_*}=0},
\ee
with 
\be
\label{eq:psic}
\psi_c = -\dfrac{2}{A'}\left(u_c'+\dfrac{\tilde{G}'}{u_c A}\sqrt{\dfrac{A}{p_\varphi^2} + u_c^2A}\right),
\ee
where
\be
\tilde{G} = G_S S + G_{S_*} S_*,
\ee
and the prime indicates derivatives with respect to $r$. The variables  $(u,u_c,p_\varphi)$ that 
appear in the definition of $r_\Omega$ (after having set $p_{r_*}\to 0$ as indicated in 
Eq.~\eqref{eq:rOmg}) are evaluated along the EOB dynamics. The definition~\eqref{eq:rOmg} 
is such that, during the adiabatic circular inspiral, one has a usual looking Kepler's law:
\be
1=\Omega^2 r_\Omega^3.
\ee
In the nonspinning limit, Eq.~\eqref{eq:rOmg} reduces to the definition of $r_\omega$ given
around Eq.~(19) in Ref.~\cite{Damour:2012ky}.
%For instance, when the dynamics is ruled by the nonspinning test-mass Kerr Hamiltonian, 
%the above definition of $r_\Omega$ reduces to
%\be
%r_\Omega^{\rm Kerr}=(r^{3/2}+a)^{2/3}.
%\ee
%In the radiation reaction, we use $x=v_\varphi^2$ as argument. Recall
%split of treatment of the odd and even-parity modes. Plus spin square
%terms in the waveform.

%Resummation: we follow Taracchini/Pan with the defactorization of the  $m$-odd modes.
%$f_{\ell m}$ for the spin part.
%We add spin square terms in some multipoles.

The azimuthal component of radiation reaction is identified with the total 
mechanical angular momentum loss given by
\be
\label{eq:Fvarphi}
{\cal F}_\varphi = {\cal F}^{\infty}_\varphi + {\cal F}^{H}_\varphi=- \dot{J}^{\infty} -\dot{J}_1^H - \dot{J}_2^H,
\ee
where the angular momentum flux at infinity $\dot{J}^\infty$ is resummed according 
to the multipolar waveform resummation introduced in~\cite{Damour:2008gu} 
for nonspinning  binaries and extended in~\cite{Pan:2011gk} to the 
spinning case.
The horizon flux contributions $\dot{J}^H_A$ are obtained by combining the
results of Refs.~\cite{Alvi:2001mx,Nagar:2011aa}.

More precisely, $\dot{J}^\infty$ is given by
\be
\label{eq:flux}
\dot{J}^{\infty}=\frac{\dot{E}^{\infty}}{\Omega}=\dfrac{1}{8\pi}\sum_{\ell=2}^8\sum_{m=1}^{\ell}m^2\Omega|
R h_{\ell m}(x)|^2, 
\ee
where $R$ denotes the distance from the source. The multipoles $h_\lm(x)$ 
are written in factorized form as
\be
\label{eq:h_lm}
%h_{\ell m}^{\rm insplunge} = h_\lm^{(N,\epsilon)}(\v)  S_{\rm eff}^{(\epsilon)}\hat{h}_\lm^{\rm tail}(y) \left[\rho_\lm(\v^2)\right]^\ell \hat{h}_\lm ^{\rm NQC},
h_{\lm}(x) = h_\lm^{(N,\epsilon)}(\v)\hat{h}_\lm^{\rm tail}(y)\hat{S}^{(\epsilon)}_{\rm eff} f_\lm(\v,S_1,S_2) \hat{h}_\lm ^{\rm NQC},
\ee
where we indicated the (main) arguments used in several factors of the waveform 
[in particular $y\equiv (H\Omega)^{2/3}$].
Note that in our EOB/NR comparisons below we shall work with a 
Zerilli-normalized multipolar 
waveform $\Psi_{\ell m}=(R/M)h_\lm/\sqrt{(\ell+2)(\ell+1)\ell(\ell-1)}$. 
Here $\epsilon=0,1$ is the parity of the considered multipole 
(i.e. the parity of $\ell+m$) . The first  factor,  $h_\lm^{(N,\epsilon)}$, 
is the Newtonian waveform,
\be
h_\lm^{(N,\epsilon)}(x)=\dfrac{\nu}R n_\lm^{(\epsilon)}c_{\ell+\epsilon}\v^{\ell + \epsilon}Y^{\ell-\epsilon,-m}\left(\dfrac{\pi}{2},\varphi_{\rm orb}\right),
\ee
where  $Y^{\lm}(\theta,\varphi)$ are scalar spherical harmonics, $\varphi_{\rm orb}$ the
orbital phase, and where the functions $n_\lm^{(\epsilon)}$ and $c_{\ell+\epsilon}$ 
are given in Eqs.~(5), (6) and (7) of Ref.~\cite{Damour:2008gu}.
Note in particular that
\be
\label{eq:cle}
c_{\ell+\epsilon}=X_2^{\ell+\epsilon-1}+(-)^m X_1^{\ell + \epsilon-1}.
\ee
The second factor is the effect of tails~\cite{Damour:2007xr,Damour:2007yf,Damour:2008gu}:
\be 
\label{htail}
\hat{h}^{\rm  tail}_\lm(y) \equiv T_\lm(y) e^{\ii\delta_\lm (y)},
\ee 
where the residual phase corrections $\delta_\lm(y)$ are given as in Ref.~\cite{Damour:2012ky},
without adding possible spin-dependent corrections in this exploratory study. The third factor, 
$\hat{S}_{\rm eff}^{(\epsilon)}$, is a parity-dependent source term defined 
as $\hat{S}_{\rm eff}^{(0)}=\hat{H}_{\rm eff}$ and $\hat{S}_{\rm eff}^{(1)}=p_\varphi/(r_\Omega v_\varphi)$.
The fourth factor $f_\lm(v_\varphi,S_1,S_2)$ is taken, when $m$ is even, as $f_\lm=\rho_\lm^\ell$
with 
\be
\rho_{\lm} = \rho_{\lm}^{\rm orb} + \rho_{\lm}^{S} .
\ee
Here, we use the orbital contributions $\rho_{\lm}^{\rm orb}$ at the $3^{+2}$PN accuracy as 
given in Ref.~\cite{Damour:2012ky}. The spin-dependent contributions $\rho^S_{\lm}$
are mostly taken from Ref.~\cite{Pan:2010hz} (modulo some new, additional contributions
mentioned below):
\begin{align}
\rho_{22}^S &= c_{\rm SO}^{\rm LO} v^3 + c_{\rm SS}^{LO} v^4 + c_{\rm SO}^{\rm NLO} v^5,\\
\rho_{32}^S &= -\dfrac{4\nu}{3(3\nu-1)}\chi_S v, \\
\rho_{44}^S &= -\dfrac{1}{15(1-3\nu)}\nonumber\\
           &\times \left[(42\nu^2-41\nu+10)\chi_S + (10-39\nu)\delta m\chi_A\right]v^3,\\
\rho_{42}^S &= -\dfrac{1}{15(1-3\nu)}\nonumber\\
           &\times\left[(78\nu^2-59\nu+10)\chi_S + (10-21\nu)\delta m\chi_A\right]v^3.
\end{align}
The $\rho_{22}^S$ term has two spin-orbit contributions (at LO and NLO) and
a LO spin-spin one~\footnote{There are no explicit spin-quadratic contributions to multipole
moments. This spin-spin term comes from the indirect effect of the spin-spin Hamiltonian 
contribution~\cite{Buonanno:2012rv}.} involving the square of the effective Kerr parameter $a=a_1+a_2=(S+S_*)/M$.
Explicitly we have
\begin{align}
c_{\rm SO}^{\rm LO}    &= -\dfrac{2}{3}\left[\chi_S(1-\nu)+\chi_A\delta m\right],\\
c_{\rm SO}^{\rm NLO}   &=   \left(-\dfrac{34}{21} + \dfrac{49}{18}\nu + \dfrac{209}{126}\nu^2\right)\chi_S \nonumber\\
                   &+ \left(-\dfrac{34}{21} - \dfrac{19}{42}\nu\right)\delta m\chi_A, \\
c_{\rm SS}^{\rm LO}    & = \dfrac{1}{2}a^2,
\end{align}
where we used the notation
\begin{align}
 \chi_A = \dfrac{1}{2}\left(\chi_1-\chi_2\right),\qquad \chi_S = \dfrac{1}{2}\left(\chi_1+\chi_2\right),
\end{align}
\be
\delta m  = X_1-X_2.
\ee
The NLO, ${\cal O}(v^5)$ spin-orbit contribution to $\rho_{22}^S$ is new. 
It was deduced by us (together with the ${\cal O}(v^3)$ contribution to $\tilde{f}_{21}^S$ discussed below) 
from the decomposition in partial multipoles (which was kindly provided to
us by Guillaume Faye) of the total energy flux given in Eq.~(5) of 
the 2010 Erratum~\cite{PhysRevD.81.089901} of Ref.~\cite{Blanchet:2006gy}.
On the other hand, when $m$ is odd, we use (essentially as in Ref.~\cite{Taracchini:2012ig}) an additive
expression for $f_\lm(v_\varphi,S_1,S_2)$, which defactorizes the singular factor $\delta m$ (which vanishes
in the equal-mass limit) present in the Newtonian prefactor $[c_{\ell+\epsilon}]_{\epsilon=1}\propto \delta m$, 
Eq.~\eqref{eq:cle}:
\be
\delta m f_\lm(v_\varphi,S_1,S_2) =\delta m \left(\rho_\lm^{\rm orb}\right)^\ell+ \tilde{f}_\lm^S,
\ee
where the $\delta m$-rescaled spin contributions $\tilde{f}^S_\lm$ are\footnote{Beware, however, of a misprint in Eq.~(A15c) of~\cite{Taracchini:2012ig}, 
giving $f_{31}$, with respect to the (corrected) results of~\cite{Pan:2010hz}.}~\cite{Taracchini:2012ig}
\begin{align}
\tilde{f}_{21}^S &= -\dfrac{3}{2}(\delta m\chi_S + \chi_A) v \nonumber\\
        & + \bigg[  \left(\dfrac{61}{12} + \dfrac{79}{84} \nu\right)\delta m\chi_S + \left(\dfrac{61}{12} + \dfrac{131}{84}\nu\right)\chi_A \bigg]v^3,\\ 
\tilde{f}_{33}^S &= -\left[ \delta m\chi_S\left(2 -  \dfrac{5}{2}\nu\right) + \chi_A\left(2 - \dfrac{19}{2}\nu\right) \right]v^3,\\    
\tilde{f}_{31}^S &= -\left[ \delta m \chi_S\left(2 - \dfrac{13}{2}\nu\right) + \chi_A\left(2 - \dfrac{11}{2}\nu\right) \right]v^3,\\       
\tilde{f}_{43}^S &= -5\dfrac{\nu}{2(2\nu-1)}(\delta m\chi_S - \chi_A)v,\\ 
\tilde{f}_{41}^S &=  \tilde{f}_{43}^S.
\end{align}
Here the ${\cal O}(v^3)$ term in $\tilde{f}_{21}^S$ is new (see discussion above of 
the NLO contribution to $\rho_{22}^S$). Finally the NQC multipolar factor 
$\hat{h}^{\rm NQC}_{\lm}$ in Eqs.~\eqref{eq:h_lm} depends on 4 real 
parameters, 2 for the amplitude, $a_i^{\ell m}$, $i=1,2$, and 2 for the phase 
$b_i^{\lm}$, $i=1,2$ and reads
\be
\label{eq:hNQC}
\hat{h}_\lm^{\rm NQC} = \left(1 + \sum_{j=1}^{2} a_j^{\lm} n_j \right)\exp\left(i\sum_{j=1}^{2}b_j^{\lm} n'_{j}\right),
\ee
where the $n_i$'s factors are chosen here to be
\begin{subequations}
\label{eq:allnqc}
\begin{align}
\label{eq:n1_nqc}
n_1 &= \left(\dfrac{p_{r_*}}{r\Omega}\right)^2\\
\label{eq:ddotr}
n_2 &= \dfrac{(\ddot{r})^{(0)}}{r\Omega^2},\\
n'_1 &= \dfrac{p_{r_*}}{r\Omega},\\
n'_2 &= p_{r_*} r\Omega=n_1' (r\Omega)^2.
\end{align}
\end{subequations}
Here, the superscript $(0)$ on the right-hand side of 
the definition of $n_2$ means that the second time derivative of 
$r$ is evaluated along the conservative dynamics 
(i.e. neglecting the contributions proportional to $\F$, see 
Appendix of~\cite{Damour:2012ky} for a discussion).
For simplicity in the present work we include NQC factors
only for $\ell=m=2$ (because of the complete mass and spin
symmetry the next multipoles $(2,1),(3,3)$ exactly vanish).
 
\subsection{Horizon flux contributions}
The horizon flux contribution ${\cal F}_{\varphi}^H$ to the azimuthal radiation 
reaction force $\F_\varphi$, Eq.~\eqref{eq:Fvarphi}, is written as
\be
{\cal F}_{\varphi}^H = -\dfrac{32}{5}\nu^2 \Omega^5 r_\Omega^4 \left(\hat{\dot{J}}_1^H+\hat{\dot{J}}_2^H\right),
\ee
where
\begin{align}
\hat{\dot{J}}^H_1 &= \dfrac{v_\varphi^5}{4}X_1^3(1+3\chi_1^2)\nonumber\\
                 &\times \left[-\chi_1 + 2 \hat{F}_{22}^{(H,0)}\left(1+\sqrt{1-\chi_1^2}\right)X_1 v_\varphi^3\right]
\end{align}
is the result derived in Ref.~\cite{Alvi:2001mx} modified by including the (resummed, hybridized) 
PN amplification factor $\hat{F}_{22}^{(H,0)}$ given by Eq.~(35) of Ref.~\cite{Nagar:2011aa}. We leave 
to future work the inclusion of the PN corrections to the spin-odd leading order term, $\propto v_\varphi^5$ 
(see Refs.~\cite{Taylor:2008xy,Taracchini:2013wfa}).

\subsection{Ringdown modelization}
Contrary to previous EOB work, the ringdown is here modeled using the new
analytical representation recently introduced in Ref.~\cite{Damour:2014yha}.
Let us recall that the EOB waveform is made of the juxtaposition of two
distinct waveforms: the inspiral-plus-plunge waveform on the EOB  time 
interval $-\infty<t_{\rm EOB}<t_{\rm match}^{\rm EOB}$ and the ringdown waveform for 
$t_{\rm EOB}> t_{\rm match}^{\rm EOB}$.
In the notation used below the matching time will be
\be
t_{\rm match}^{\rm EOB}=t^{\rm EOB}_{\rm NQC}
\ee
which corresponds, on the NR time axis, to 
the instant $t_{\rm match}^{\rm NR}=t^{NR}_{A^{\rm max}_{22}}+2M$, where 
$A_{22}\equiv |h_{22}^{\rm NR}|$ is the amplitude of the NR quadrupolar
waveform $h_{22}^{\rm NR}$.

%====================================================
\section{NR completion of the new spinning EOB model}

\subsection{Alignment of the EOB and NR time axes and determination of NQC corrections}

In early (nonspinning) EOB work the correspondence between the EOB and NR time axes was
defined by identifying the peak of the orbital EOB frequency $\teobLR$  with the peak of the 
$\ell=m=2$ NR amplitude $\tAmax$ . However, later 
work~\cite{Pan:2011gk,Taracchini:2012ig,Damour:2012ky}  
introduced as a free parameter a time-lag between the peak of the 
orbital $\teobLR$ and $\tAmax$.
Here we shall make use of this flexibility but we shall parametrize
it in a different way by using as main ``anchor point'' on the EOB
time axis the peak $\teobpo$ of the {\it pure orbital frequency}
\be
\label{eq:Omg_Orb}
\Omega_{\rm orb} \equiv \dfrac{M}{H}\dfrac{\de H_{\rm orb}^{\rm eff}}{\de L}= \dfrac{p_\varphi u_c^2 A}{H \hat{H}_{\rm orb}^{\rm eff}}.
\ee
The motivation for this choice is the following: while, in the test mass limit, 
$\tAmax$ differed from $\teobLR$ by an amount which was getting as large as several 
tens of $M$ for large and positive spin~\cite{Barausse:2011kb}, it was found that (in the
test-mass limit) $\tAmax$ differed from $\teobpo$ only by a few $M$'s for most spin 
values~\cite{Harms:2014dqa}. In the present work we parametrize this flexibility 
by introducing the quantity $\Delta t_{\rm NQC}$ defined so that
\be
\label{eq:DTnqc}
t^{\rm EOB}_{\rm NQC}\equiv t_{\Omega_{\rm orb}}^{\rm peak}-\Delta t_{\rm NQC} .
\ee
In addition we define the correspondence between the EOB and NR time axes 
by requiring that
\be
\teobNQC \leftrightarrow \tnrextr,
\ee
where we choose the NR extraction point $t^{\rm NR}_{\rm extr}$ to be $2M$  on the right of 
the peak of the NR $h_{22}$ waveform, i.e.
\be
\tnrextr\equiv \tAmax + 2M.
\ee 
This choice means that the peak on $h_{22}$ on the NR time axis 
corresponds to
\be
\tAmax \leftrightarrow \teobpo - 2M - \Delta t_{\rm NQC}.
\ee
Analogously to Ref.~\cite{Damour:2012ky} (mutatis mutandis),
the degree of osculation between the  EOB and NR waveforms 
at $t_{\rm EOB}^{\rm NQC} \leftrightarrow t_{\rm NR}^{\rm extr}$ 
is defined by imposing the following four conditions
\begin{subequations}
\label{eq:C2_cond}
\begin{align}
\label{comp_1}
A_{\lm}^{\rm EOB}(\teobNQC)             & = A_{\lm}^{\rm NR}(\tnrextr),\\ 
\label{comp_2}
\dot{A}_{\lm}^{\rm EOB}(\teobNQC)       &= \dot{A}_{\lm}^{\rm NR}(\tnrextr),\\ 
\label{comp_4}
\omega_{\lm}^{\rm EOB}(\teobNQC)        &= \omega_{\lm}^{\rm NR}(\tnrextr),\\ 
\label{comp_5}
\dot{\omega}_{\lm}^{\rm EOB}(\teobNQC)  &= \dot{\omega}_{\lm}^{\rm NR}(\tnrextr),
\end{align}
\end{subequations}
which yield both a $2\times 2$ linear system to be solved to
obtain the $a_i^\lm$'s, and, separately, a $2\times 2$ 
linear system to be solved for the $b_i^\lm$'s.
Note that the values of the  $a_j^\lm$'s affect the modulus of the inspiral-plus-plunge
waveform, which then affects the computation of the radiation reaction force (through
the  angular momentum flux). In turn, this modifies the EOB dynamics itself,
and, consequently, the determination of the $(a_j^\lm,b_j^\lm)$'s.
This means that one must bootstrap, by iteration, the determination
of the $(a_j^\lm,b_j^\lm)$'s until convergence (say at the third decimal digit)
is reached. This typically requires three iterations.
%============================
% FIG.1: Nonspinning case
%============================
\begin{figure}[t]
\begin{center}
 \includegraphics[width=0.48\textwidth]{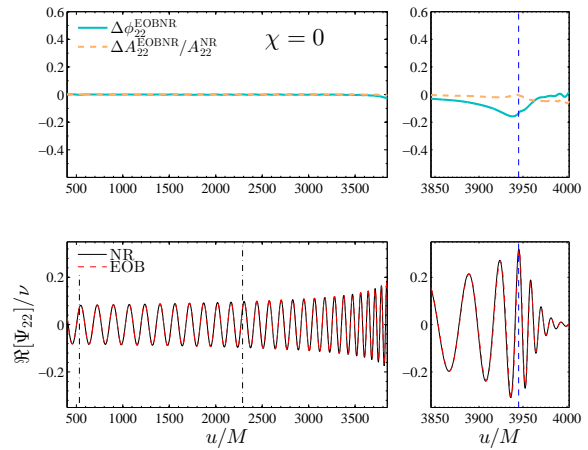}
    \caption{ \label{fig:chi0}Performance of the recalibrated nonspinning limit of our new SEOB model. The vertical
    dashed line indicates the NR NQC extraction point and matching time $t_{\rm extr}^{\rm NR}$. The two dash-dotted vertical
    lines during inspiral indicate the alignment time window corresponding to $\ell=m=2$ 
    GW frequencies $[M\omega_L,M\omega_R]=[0.035,0.045]$. See text for details.} 
\end{center}
\end{figure}
%-----------------

%===========================================
% Table for the NQC parameters actually used
%===========================================
\begin{table}[t]
  \caption{\label{tab:nqc_ab}Parameters of the model: effective NNNLO spin-orbit coupling,  $c_3$,
   NR/EOB time-shift $\Delta t_{\rm NQC}$ and next-to-quasi-circular $(a_i^\lm,b_i^\lm)$ parameters for the 
   sixteen values of the spin considered. The NQC parameters are obtained by imposing osculation 
   conditions between EOB and NR waveform amplitudes and frequencies around merger.}  
  \centering  
  \begin{ruledtabular}  
  \begin{tabular}{ccccccc}        
    \hline
    $\chi$ & $c_3$ & $\Delta t_{\rm NQC}$ & $a_1^{22}$ & $a_2^{22}$ & $b_1^{22}$ & $b_2^{22}$  \\
    \hline    
    \hline
    $-0.94905 $ & 92.5    &         $\;\;\;1$  & $\;\;\;0.143$    & 0.337   & 0.163   & 3.489 \\
    $-0.8996$   & 89      &         $\;\;\;1$  & $\;\;\;0.139$    & 0.364   & 0.165   & 3.362 \\
    $-0.7998$   & 84      &         $\;\;\;1$  & $\;\;\;0.135$    & 0.442   & 0.161   & 2.895 \\
    $-0.5999$   & 73      &         $\;\;\;1$  & $\;\;\;0.129$   & $0.614$ & $0.156$ & 2.090\\
    $-0.43756$  & 67      &         $\;\;\;1$  & $\;\;\;0.128$   & $0.786$ & 0.150 & 1.583\\
    $-0.2000$   & 59      &         $\;\;\;1$  & $\;\;\;0.126$   & $1.032$ & 0.147 & 1.018 \\
    $ 0.0 $     & $\dots$ &         $\;\;\;1$  & $-0.075$        & $1.496$ & 0.148 & 0.914 \\
    $+0.2000$   & 26      &         $\;\;\;1$  & $ 0.106$        & $1.574$ & 0.134 & 0.564\\
    $+0.436554$ & 17      &         $\;\;\;1$  & $-0.023$        & $1.862$ & 0.086 & 0.263\\
    $+0.6000$   & 16.5    &         $\;\;\;1$  & $-0.129$        & $1.598$  & 0.083 & 1.016\\
    $+0.7999$   & 8.5     &         $\;\;\;1$  & $-0.233$        & $0.632$  & 0.572 & 7.947\\
    $+0.8498$   & 5.5     &         $\;\;\;1$  & $-0.313$        & $-0.244$ & 0.809 & 16.061\\
    $+0.8997$   & 5.5     &         $-1$       & $\;\;\;0.079$   &  0.214 & 0.768 & 12.607 \\
    $+0.9496$   & 4.5     &         $-3$       & $\;\;\;0.432$   &  0.338 & 0.678 & 10.900 \\
    $+0.9695$   & 4.5     &         $-4$       & $\;\;\;0.612$   &  0.562 & 0.525 & 8.328 \\
    $+0.9794$   & 3.5     &         $-4$       & $\;\;\;0.608$   &  0.214 & 0.638 & 12.058 \\
    \hline
  \end{tabular}
  \end{ruledtabular}  
\end{table}

%===================================
% Fig.2: phasing in presence of spin
%===================================
\begin{figure*}[t]
\begin{center}
 \includegraphics[width=0.45\textwidth]{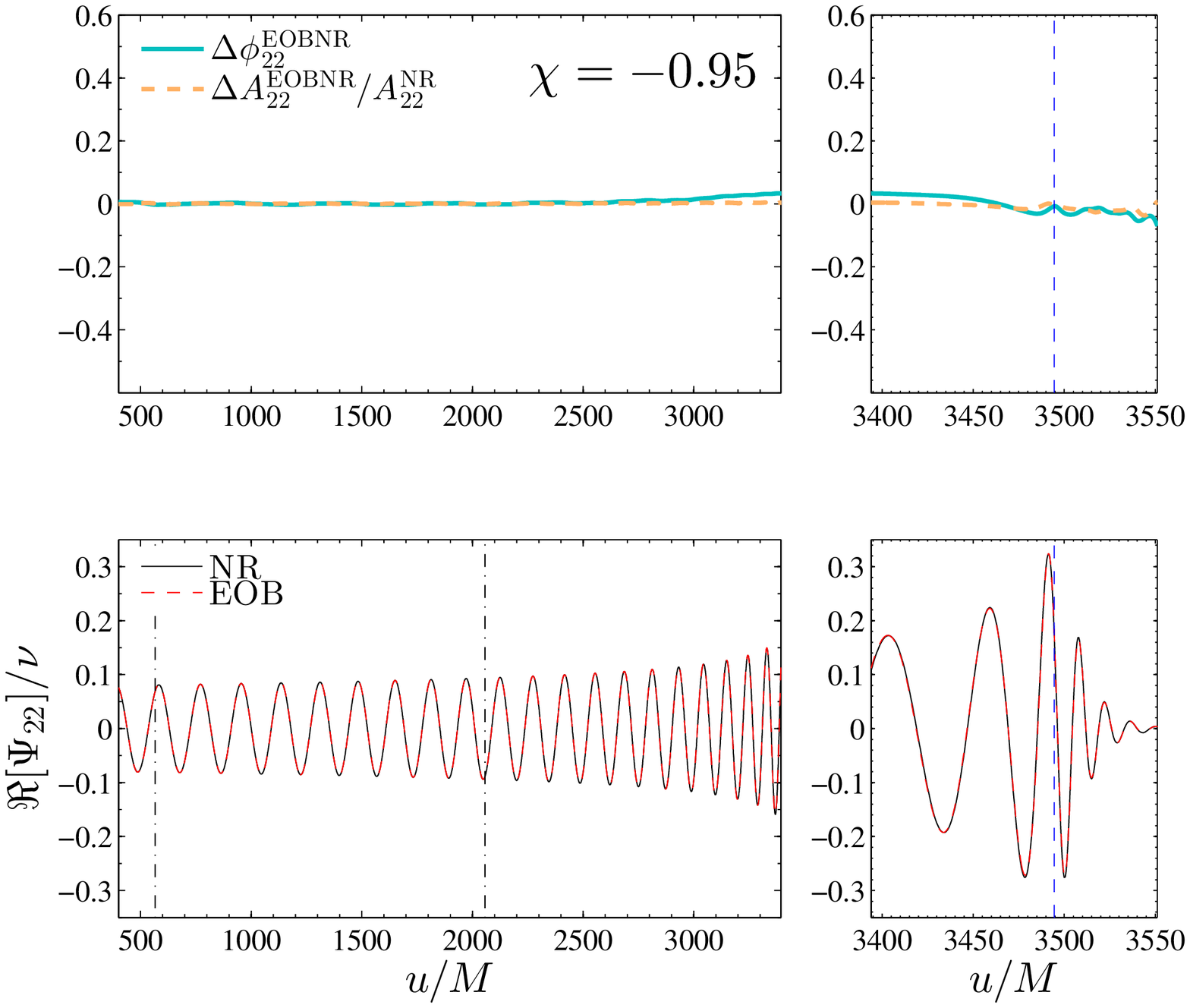}
 \includegraphics[width=0.45\textwidth]{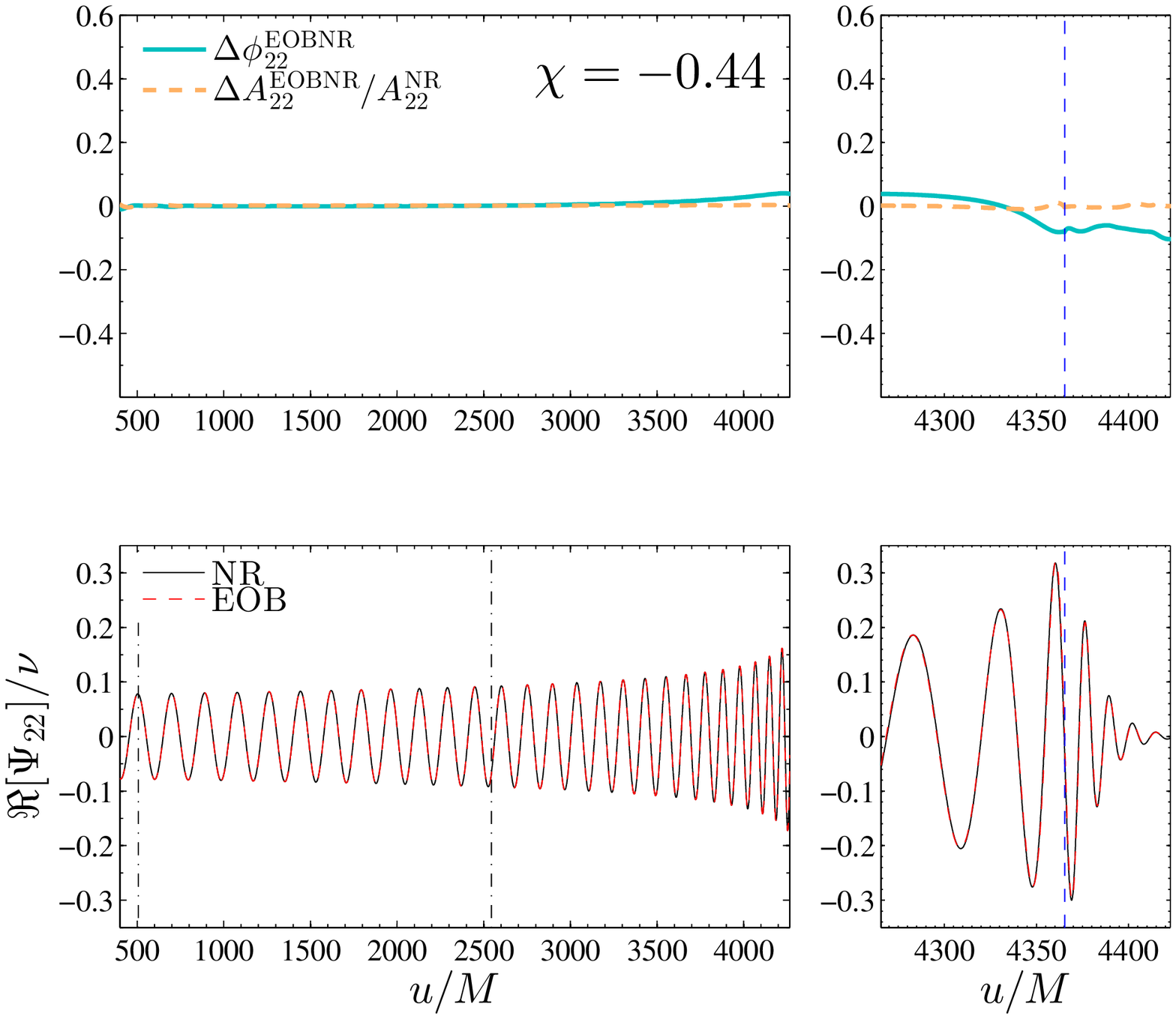}\\
 \includegraphics[width=0.45\textwidth]{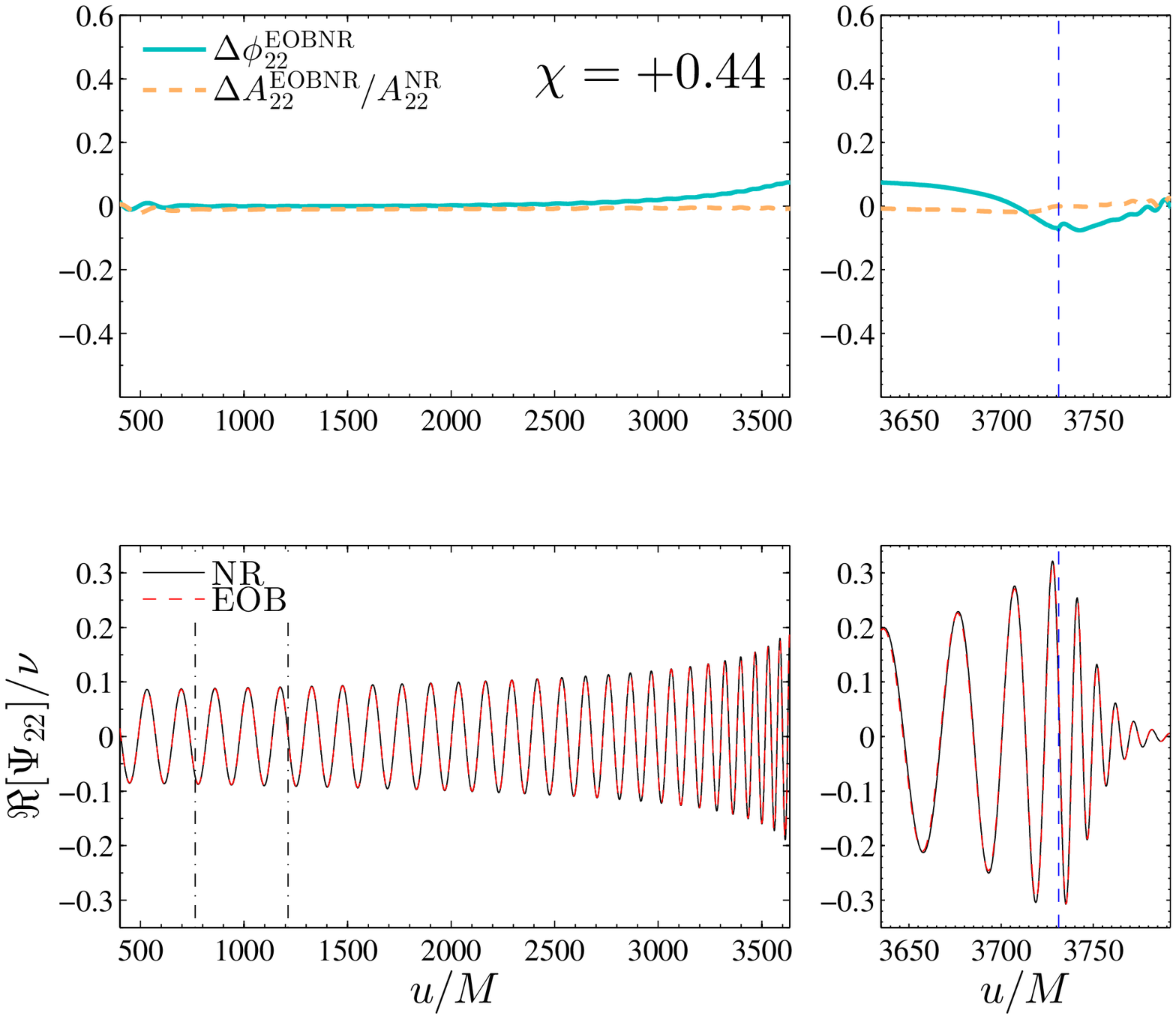}
 \includegraphics[width=0.45\textwidth]{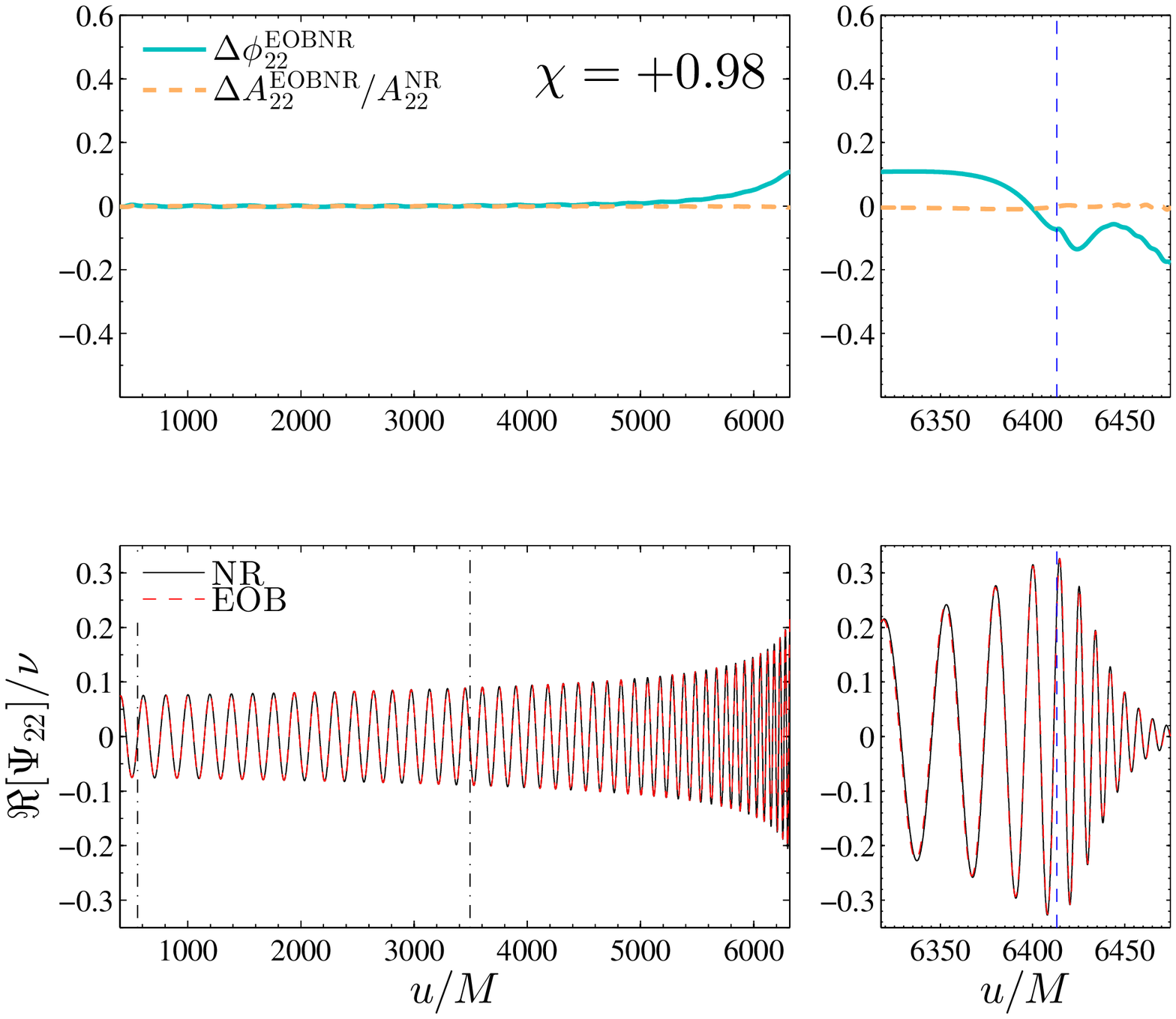}\\
    \caption{ \label{fig:chi_all}SEOB/NR $\ell=m=2$ (Zerilli-normalized) waveform comparison 
     when $\chi\neq 0$. Waves are aligned at early times for all spins. 
     The dashed vertical line indicates the NR NQC extraction point and matching time $t^{\rm NR}_{\rm extr}$.
     The two dash-dotted vertical lines during inspiral indicate the alignment time window. 
     See text for details.} 
\end{center}
\end{figure*}
%-----------------

\subsection{Analytical freedom of the model and its NR calibration}
In summary, the adjustable parameters of the present spinning EOB model 
are: $\left(a_6^c,c_3,\Delta t_{\rm NQC}\right)$.
As said above the radiation reaction used in the present EOB model differs
from the one of Ref.~\cite{Damour:2012ky} by setting to zero the radial 
component $\F_r=0$. This change in radiation reaction obliged us to (marginally) 
recalibrate the nonspinning model of Ref.~\cite{Damour:2012ky}, which means, in the 
presently considered equal-mass case, the values of $a_6^c$ and $\Delta t_{\rm NQC}$. 
In doing so, we could also simplify the NQC factor of Ref.~\cite{Damour:2012ky},
reducing the NQC parameters from the six used there (3 for the amplitude and 3 for the phase) 
to the four we use here. We found that the  values
\begin{align}
a_6^c                    & = -129,\\
\Delta t_{\rm NQC}(\chi=0) & =  1M,
\end{align}
give a EOBNR phase agreement comparable to the one of ~\cite{Damour:2012ky},
as illustrated in Fig.~\ref{fig:chi0}. The NR/EOB alignment time window 
is indicated by the two dashed vertical lines and corresponds to $\ell=m=2$ 
GW frequencies $[M\omega_L,M\omega_R]=[0.035,0.045]$ (see below). Note that 
the value $\Delta t_{\rm NQC}=1M$ corresponds to a time lag between
$\tAmax$ and $\teobpo$ equal to $\teobpo=\tAmax+3M$.

In the (equal-mass) spinning case, we keep the value $a_6^c = -129$ and adjust, for each
value of $\chi$, $c_3$ and $\Delta t_{\rm NQC}$ so as to minimize the phase difference
between the EOB and NR $\ell=m=2$ waveform. We find that the value $\Delta t_{\rm NQC}=1M$ 
is a good compromise when 
$\chi \leq 0.85$. For the other values one discovers that values 
$\Delta t_{\rm NQC}< 1M$ are needed to improve the phasing agreement around merger.
The values of $(c_3,\Delta t_{\rm NQC})$ we used are listed in Table~\ref{tab:nqc_ab},
together with the corresponding $(a_i^{22},b_i^{22})$'s.
The values of $c_3$ leading to a good phase agreement were found to depend (in a roughly
piecewise linear manner) on $\chi$. Such a $\chi$ dependence of the effective NNNLO spin-orbit 
parameter $c_3$ can be intuitively understood if one remembers that we have parametrized 
NNNLO spin-orbit effects by a purely $r$-dependent term 
$\nu c'_3/r^3$ while we could have also included, at the 
same PN order, terms of order $\nu c_3''p_{r_*}^2/r^2$. In other words the single, 
effective, value of $c_3$ we use can be roughly considered as being of order 
$c_3\approx c'_3 + \langle r p{_{r_*}}^2\rangle c_3''$ where $\langle r p{_{r_*}}^2\rangle$
is some average value of  $r p_{r_*}^2$ during the plunge.
When $\chi>0$ the plunge phase becomes shorter and shorter as $\chi$ increases, while 
when $\chi<0$ the plunge phase becomes more prominent as $\chi\to -1$. Correspondingly 
we expect that the average value $\langle r p_{r_*}^2\rangle$ will depend on $\chi$ and 
take significantly larger values for $\chi \to -1$. These considerations might explain
why we had to use a $\chi$-dependent value (increasing when $\chi\to -1$) for the 
effective NNNLO parameter $c_3$.
We leave to future work a study including more general ways of parametrizing NNNLO
effects, possibly including ${\bf p}^2$-dependent terms in $(G_S,G_{S_*})$.

These time-domain comparisons are done by suitably determining a relative time and phase shift 
between the two phases $\phi_{22}^{\rm NR}(t^{\rm NR})$ and  $\phi_{22}^{\rm EOB}(t^{\rm EOB})$.
These shifts are estimated by minimizing  the time integral of the square of the phase difference on 
a time interval corresponding to a given frequency interval $[M\omega_L,M\omega_R]$.
Following Refs.~\cite{Boyle:2008ge,Pan:2011gk,Damour:2012ky} , we usually perform this waveform 
alignment  on the long inspiral phase. More precisely, we take $M\omega_R=0.041$ for all 
waveforms, while $M\omega_L$ ranges between 0.031 and 0.033 for all spin values 
except for $\chi=+0.44$, where we take $M\omega_L= 0.0386$.
Figure~\ref{fig:chi_all} illustrates the EOB/NR agreement  for the $\ell=m=2$ waveform (both in phase
and amplitude) for the representative values of $\chi=(-0.95,-0.44,+0.44,+0.98)$.
The dashed vertical line on the plots indicates the NR NQC extraction $t^{\rm NR}_{\rm extr}$.
The alignment time window corresponding to $[M\omega_L,M\omega_R]$ is indicated by the
two dash-dotted vertical lines in the inspiral phase.

The amplitude agreement (see dashed, orange online, lines) is excellent all over, including during 
the ringdown. The phasing disagreement is always remarkably small and constant during most of the
inspiral. It remains globally within $\pm 0.15$~rad.
We note in this respect that the recent SEOB model of~\cite{Taracchini:2013rva} exhibits a phase 
disagreement of 0.6~rad with the same SXS $\chi=+0.98$ waveform (after tuning more parameters 
and using and alignment around merger), and that for a similar (though older) $\chi=+0.97$ simulation 
the accumulated NR phase error was estimated in Ref.~\cite{Lovelace:2011nu} to be of the order 
of 0.9 rad up to merger and ringdown.

\section{Conclusions}
We have presented a new (nonprecessing) spinning EOB model. The most novel feature of our scheme
is the use (for equatorial dynamics) of the concept of {\it centrifugal radius} $r_c$. We showed
the interest of this concept for the test-mass Kerr Hamiltonian and used its comparable-mass-case
generalization to define a new way of blending the spin-deformation with the mass-ratio-deformation.
In addition, we have used a recently proposed new way to attach the ringdown waveform,
based on a new (NR-fitted) way of parametrizing the ringdown signal.

In the present exploratory study, we considered a minimal version of our new SEOB model 
having only {\it three} ajustable
parameters: $a_6^c$, $\Delta t_{\rm NQC}$ and $c_3$. The nonspinning limit of this model differs
(from the analytical point of view) from our previous nonspinning model~\cite{Damour:2012ky} 
essentially in having set the radial component of the radiation reaction $\F_r$ to zero. This led us to 
recalibrate the value of $a_6^c$ against nonspinning SXS NR waveforms, which yielded 
$a_6^c=-129$ (instead of our previous preferred value $a_6^c=-101.876$). Such a calibration 
(together with a modified choice of $\Delta t_{\rm NQC}$) yields a $\ell=m=2$ waveform whose 
phasing agrees within $\pm 0.08$ rad with the $q=1$, nonspinning, SXS NR waveform.

Then, by calibrating (for each value of the spin $\chi$) the single, effective, 
next-to-next-to-next-to-leading-order spin-orbit parameter $c_3$ we were able 
to obtain a good EOB/NR phasing agreement over the full time span of each of the fifteen, 
presently catalogued spinning SXS waveforms. The range of spin values spanned by these
waveforms is $-0.95\leq \chi\leq 0.98$. The longest waveforms have about 50 GW cycles up to merger.
In the case of $\chi=0.98$ the accumulated EOB/NR phasing disagreement over the 50 GW cycles
before merger (corresponding to $\Delta t\sim 6420M$) is of the order of $0.1$~rad.

The present study leaves room for many possible improvements and further investigations such as:
(i) including the recently determined 4PN Hamiltonian~\cite{Bini:2013zaa,Damour:2014jta} as
well as the gravitational-self-force knowledge of the spin-orbit coupling~\cite{Dolan:2013roa,Bini:2014ica};
(ii) exploring the role of the various ways of gauge-fixing the spin-orbit couplings so as to
eliminate the $\chi$-dependence of the effective spin-orbit parameter $c_3$ or at least to 
introduce some additional analytical flexibility (e.g., of the type $c''_3$ mentioned above)
leading to a smooth, and therefore fittable, $\chi$-dependence.

\acknowledgments
We are grateful to Guillaume Faye for kindly providing us the multipolar decomposition of the spin-orbit
contribution to the energy flux of Ref.~\cite{Blanchet:2006gy}. AN thanks Tony Chu, Geoffrey Lovelace, 
Yi Pan and Bela Szil\`agyi for useful discussions during the 2013 NRDA meeting, where a preliminary 
version of this work was presented. We thank ICRANet for partial support.

\bibliography{refs20140626.bib}{}     

\end{document}